\documentclass[aps,prb,fleqn,12pt,showpacs,superscriptaddress]{revtex4}
\usepackage{epsfig,color}
\usepackage{graphicx}
\usepackage{pdfpages}
\usepackage{braket}
\usepackage{amssymb,amsmath}
\usepackage{amsfonts} 
\usepackage{hyperref} 

\begin{document}
\title{Correlated motion of particle-hole excitations \\ 
across the renormalized spin-orbit gap in $\rm Sr_2 Ir O_4$}
\author{Shubhajyoti Mohapatra}
\affiliation{Department of Physics, Indian Institute of Technology, Kanpur - 208016, India}
\author{Avinash Singh}
\email{avinas@iitk.ac.in}
\affiliation{Department of Physics, Indian Institute of Technology, Kanpur - 208016, India}
\date{\today} 
\begin{abstract}
The high-energy collective modes of particle-hole excitations across the spin-orbit gap in $\rm Sr_2IrO_4$ are investigated using the transformed Coulomb interaction terms in the pseudo-spin-orbital basis constituted by the $J=1/2$ and $3/2$ states arising from spin-orbit coupling. With appropriate interaction strengths and renormalized spin-orbit gap, these collective modes yield two well-defined propagating spin-orbit exciton modes, with energy scale and dispersion in excellent agreement with resonant inelastic X-ray scattering (RIXS) measurements. 
\end{abstract}
\maketitle
\newpage

\section{Introduction}
The iridium based transition-metal oxides exhibiting novel $J$=1/2 Mott insulating states have attracted considerable interest in recent years in view of their potential for hosting collective quantum states such as quantum spin liquids, topological orders, and high-temperature superconductors.\cite{krempa_AR_2014} The effective $J$=1/2 antiferromagnetic (AFM) insulating state in iridates arises from a novel interplay between crystal field, spin-orbit coupling (SOC) and intermediate Coulomb correlations. Exploration of the emerging quantum states in the iridate compounds therefore involves investigation of the correlated spin-orbital entangled electronic states and related magnetic properties. 

Among the iridium compounds, the quasi-two-dimensional (2D) square-lattice perovskite-structured iridate $\rm Sr_2IrO_4$ is of special interest as the first spin-orbit Mott insulator to be identified and because of its structural and physical similarity with $\rm La_2CuO_4$.\cite{rau_AR_2016,bertinshaw_AR_2018} It exhibits canted AFM ordering of the pseudospins below N\'{e}el temperature $T_{\rm N} \approx 240$ K. The canting of the in-plane magnetic moments tracks the staggered $\rm IrO_6$ octahedral rotations about the $c$ axis. The effectively single (pseudo) orbital ($J$=1/2) nature of this Mott insulator has motivated intensive finite doping studies aimed at inducing the superconducting state as in the cuprates.\cite{senthil_PRL_2011,kim3_SC_2014,torre_PRL_2015,kim4_NAT_2016,gretarsson_PRL_2016,chen_NATCOM_2018,
bhowal_JPCM_2018}

Technological advancements and improved energy resolution in resonant inelastic X-ray scattering (RIXS) have been instrumental in the elucidation of the pseudospin dynamics in $\rm Sr_2IrO_4$.
Recent measurements point to a partially resolved $\sim$30 meV magnon gap at the $\Gamma$ point,\cite{pincini_PRB_2017} which has been further resolved via high-resolution RIXS and inelastic neutron scattering (INS), both of which indicate another magnon gap between 2 to 3 meV at $(\pi,\pi)$.\cite{porras_PRB_2019} These low-energy features correspond to different magnon modes associated with basal-plane and out-of-plane fluctuations, indicating the presence of anisotropic spin interactions. In addition to magnon modes, RIXS experiments have also revealed a high-energy dispersive feature in the energy range 0.4-0.8 eV. Attributed to  electron-hole pair excitations across the spin-orbit gap between the $J$=1/2 and 3/2 bands, this distinctive mode is referred to as the spin-orbit exciton.\cite{kim1_PRL_2012,kim2_PRL_2012,kim_NATCOMM_2014,igarashi_PRB_2014,lu_PRB_2018}

Among the theoretical approaches, the spin-orbit exciton was identified as a bound state in the spectral function of the two-particle Green's function within the multi-orbital itinerant electron picture.\cite{igarashi_PRB_2014} However, the full dispersion was not obtained, and the original $t_{2g}$ basis was employed instead of the more natural SOC-split $J$ states with intrinsic spin-orbit gap. In another approach, the exciton dispersion was obtained in analogy with hole motion in an AFM background.\cite{kim1_PRL_2012,kim_NATCOMM_2014} However, the bare exciton dispersion was neglected, and an approach which allows for a unified description of both magnon and spin-orbit exciton on the same footing will be desirable as both excitations are observed in the same RIXS measurements. 



In this paper, we therefore plan to investigate the correlated motion of inter-orbital particle-hole excitations across the renormalized spin-orbit gap (between $J$=1/2 and $J$=3/2 sectors), along with detailed comparison with RIXS data for the spin-orbit exciton modes in $\rm Sr_2IrO_4$. Similar comparison for the magnon dispersion involving intra-orbital ($J$=1/2) particle-hole excitations has provided experimental evidence of several distinctive features associated with the rich interplay of spin-orbit coupling, Coulomb interaction, and realistic multi-orbital electronic band structure, such as (i) finite-$U$ and finite-SOC effects, (ii) mixing and coupling between the $J$=1/2 and 3/2 sectors, and (iii) Hund's-coupling-induced true magnetic anisotropy and magnon gap.\cite{iridate_one,iridate_two,igarashi_JPSJ_2014} 

The structure of the paper is as follows. After a brief account of the transformed Coulomb interaction terms in the pseudo-spin-orbital basis in Sec. II, the AFM state of the three orbital model is discussed in Sec. III. The spin-orbit gap renormalization due to the relative energy shift between the $J$=1/2 and 3/2 sectors arising from the density interaction terms is discussed in Sec. IV. The spin-orbit exciton as a resonant state formed by the correlated propagation of the inter-orbital, spin-flip, particle-hole excitation across the renormalized spin-orbit gap is investigated in Sec. V. Finally, conclusions are presented in Sec. VI.

\section{Coulomb interaction in the pseudo-spin-orbital basis}
Due to large crystal-field splitting ($\sim$3 eV) in the $\rm Ir O_6$ octahedra, the low-energy physics in $d^5$ iridates is effectively described by projecting out the empty ${\rm e_g}$ levels which are well above the $t_{\rm 2g}$ levels. Spin-orbit coupling (SOC) further splits the t$_{2g}$ states into (upper) $J$=1/2 doublet and (lower) $J$=3/2 quartet with an energy gap of $3\lambda/2$. Four of the five electrons fill the $J$=3/2 states, leaving one electron for the $J$=1/2 sector, rendering it magnetically active in the ground state. 

\begin{figure}
\vspace*{0mm}
\hspace*{0mm}
\psfig{figure=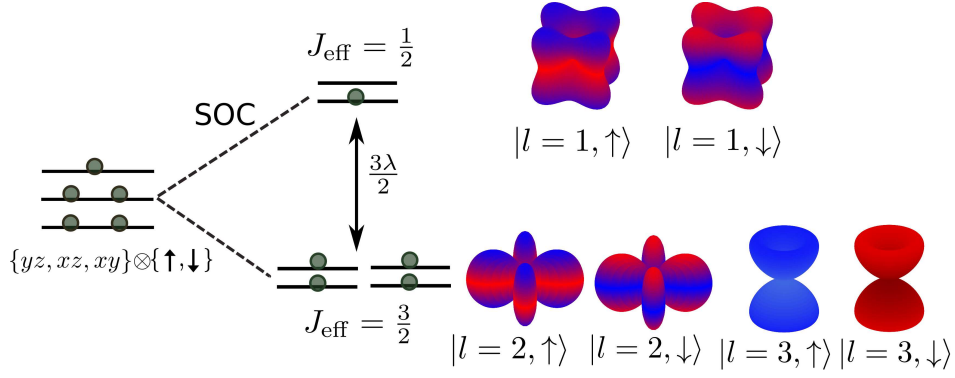,angle=0,width=125mm}
\vspace{-0mm}
\caption{The pseudo-spin-orbital energy level scheme for the three Kramers pairs along with their orbital shapes. The colors represent the weights of real spin $\uparrow$ (red) and $\downarrow$ (blue) in each pair.} 
\label{schematic}
\end{figure}

The three Kramers pairs above correspond to {\em pseudo orbitals} ($l=1,2,3$) with {\em pseudo spins} ($\tau= \uparrow,\downarrow$) each, with the $|J,m_j \rangle$ and corresponding $|l,\tau \rangle$ states having the form:
\begin{eqnarray}
\ket{l=1, \tau= \sigma} &=& \Ket{\frac{1}{2},\pm\frac{1}{2}} = \left [\Ket{yz,\bar{\sigma}} \pm i \Ket{xz,\bar{\sigma}} \pm \Ket{xy,\sigma}\right ] / \sqrt{3} \nonumber \\
\ket{l=2, \tau= \sigma} &=& \Ket{\frac{3}{2},\pm\frac{1}{2}} = \left [\Ket{yz,\bar{\sigma}} \pm i \Ket{xz,\bar{\sigma}} \mp 2 \Ket{xy,\sigma} \right ] / \sqrt{6} \nonumber \\
\ket{l=3, \tau= \bar{\sigma}} &=& \Ket{\frac{3}{2},\pm\frac{3}{2}} = \left [\Ket{yz,\sigma} \pm i \Ket{xz,\sigma}\right ] / \sqrt{2} 
\label{jmbasis}
\end{eqnarray}
where $\Ket{yz,\sigma}$, $\Ket{xz,\sigma}$, $\Ket{xy,\sigma}$ are the t$_{2g}$ states and the signs $\pm$ correspond to spins $\sigma = \uparrow/\downarrow$. The coherent superposition of different-symmetry $t_{\rm 2g}$ orbitals, with opposite spin polarization between $xz$/$yz$ and $xy$ levels implies spin-orbital entanglement, and also imparts unique extended 3D shape to the pseudo-orbitals $l=1, 2, 3$, as shown in Fig \ref{schematic}. 

Inverting the above transformation, the three real-spin-orbital basis states can be represented in terms of the pseudo-spin-orbital basis states, given below in terms of the corresponding creation operators:
\begin{equation}
\begin{pmatrix}
a_{yz \sigma}^\dagger \\ a_{xz \sigma}^\dagger \\ a_{xy \overline{\sigma}}^\dagger
\end{pmatrix} = \begin{pmatrix} \frac{1}{\sqrt{3}} & \frac{1}{\sqrt{6}} & \frac{1}{\sqrt{2}} \\
\frac{i\sigma}{\sqrt{3}} & \frac{i\sigma}{\sqrt{6}} & \frac{-i\sigma}{\sqrt{2}} \\
\frac{-\sigma}{\sqrt{3}} & \frac{\sqrt{2}\sigma}{\sqrt{3}} & 0
\end{pmatrix} \begin{pmatrix} a_{1 \tau}^\dagger \\ a_{2 \tau}^\dagger \\ a_{3 \tau}^\dagger
\end{pmatrix}
\end{equation}
where, $\sigma = \uparrow/\downarrow$ and $\tau = \overline{\sigma}$. 

We consider the on-site Coulomb interaction terms:
\begin{eqnarray}
\mathcal{H}_{\rm int} &=& U\sum_{i,\mu}{n_{i\mu\uparrow}n_{i\mu\downarrow}} + U^\prime \sum_{i,\mu < \nu,\sigma} {n_{i\mu\sigma}n_{i\nu\overline{\sigma}}} + (U^\prime - J_{\mathrm H}) \sum_{i,\mu < \nu,\sigma}{n_{i\mu\sigma}n_{i\nu\sigma}} \nonumber\\ 
&&+ 
J_{\mathrm H} \sum_{i,\mu \ne \nu} ({a_{i \mu \uparrow}^{\dagger}a_{i \nu\downarrow}^{\dagger}a_{i \mu \downarrow} a_{i \nu \uparrow}} + {a_{i \mu \uparrow}^{\dagger} a_{i \mu\downarrow}^{\dagger}a_{i \nu \downarrow} a_{i \nu \uparrow}})
\label{inter}
\end{eqnarray} 
in the real-spin-orbital basis ($\mu,\nu = yz, xz, xy$), including the intra-orbital $(U)$ and inter-orbital $(U')$ density interaction terms, the Hund's coupling term $(J_{\rm H})$, and the pair hopping term $(J_{\rm H})$. Here $a_{i\mu\sigma}^{\dagger}$ and $a_{i\mu \sigma}$ are the creation and annihilation operators for site $i$, orbital $\mu$, spin $\sigma=\uparrow ,\downarrow$, and the density operator $n_{i \mu \sigma} = a_{i \mu \sigma}^\dagger a_{i\mu \sigma}$. 

Using the transformation from the $t_{2g}$ basis to the pseudo-spin-orbital basis given above, and keeping the Hubbard, density, and Hund's coupling like interaction terms which are relevant for the present study, we obtain (for site $i$):
\begin{eqnarray}
\mathcal{H}_{\rm int} (i) &=& 
\frac{1}{2} \sum_{m, m^\prime, \tau, \tau^\prime} 
\mathcal{U}_{m m^\prime}^{\tau \tau^\prime} n_{m \tau} n_{m^\prime \tau^\prime} 
+ \left( \frac{U - U ^\prime}{3} \right) \sum_{\tau} {a_{1 \tau}^{\dagger}a_{2 \overline{\tau}}^{\dagger}a_{1 \overline{\tau}} a_{2 \tau}} \nonumber \\
& & + 
\left( \frac{U - 2J_{\rm H} - U ^\prime}{6} \right) \sum_{\tau} \left( a_{2 \tau}^{\dagger}a_{3 \overline{\tau}}^{\dagger}a_{2 \overline{\tau}} a_{3 \tau} + 2 a_{3 \tau}^{\dagger}a_{1 \overline{\tau}}^{\dagger}a_{3 \overline{\tau}} a_{1 \tau}  \right) 
\label{h_int_detail}
\end{eqnarray}
where the transformed interaction matrices $\mathcal{U}_{m m^\prime}^{\tau \tau^\prime}$ in the new basis $(m,m'=1,2,3)$:
\begin{eqnarray}
\mathcal{U}_{m m^\prime}^{\tau \tau} &=& \left( 
\begin{array}{cccc}
0  & U^\prime & U^\prime - \frac{2}{3} J_{\rm H} \\
U^\prime & 0 & U^\prime-\frac{1}{3} J_{\rm H} \\
U^\prime-\frac{2}{3} J_{\rm H} & U^\prime-\frac{1}{3} J_{\rm H} & 0
\end{array}
\right), \nonumber \\
\mathcal{U}_{m m^\prime}^{\tau \overline{\tau}} &=& \left( 
\begin{array}{cccc}
\frac{1}{3} (U + 2 U^\prime)  & \frac{1}{3} (U + 2 U^\prime - 3 J_{\rm H}) & \frac{1}{3} (U + 2 U^\prime - J_{\rm H}) \\ 
\frac{1}{3} (U + 2 U^\prime - 3 J_{\rm H}) & \frac{1}{2} (U + U^\prime) & \frac{1}{6} (U + 5 U^\prime - 4 J_{\rm H}) \\
\frac{1}{3} (U + 2 U^\prime - J_{\rm H}) & \frac{1}{6} (U + 5 U^\prime - 4 J_{\rm H}) & \frac{1}{2} (U +  U^\prime) 
\end{array} 
\right)
\label{cal_u_eqns}
\end{eqnarray}
for pseudo-spins $\tau^\prime = \tau$ and $\tau^\prime = \overline{\tau}$, where $\tau = \uparrow, \downarrow$. Similar transformation to the $J$ basis has been discussed recently, focussing only on the density interaction terms.\cite{martins_JPCM_2017}

Using the spherical symmetry condition ($U^\prime$=$U$-$2J_{\mathrm H}$), the transformed interaction Hamiltonian (\ref{h_int_detail}) simplifies to:
\begin{eqnarray}
{\mathcal H}_{\rm int}(i) &=& \left( U - \frac{4}{3} J_{\rm H} \right) n_{1 \uparrow} n_{1 \downarrow} + \left( U - J_{\rm H} \right) \left[ n_{2 \uparrow}  n_{2 \downarrow} +  n_{3 \uparrow} n_{3 \downarrow} \right] \nonumber \\
&-& \frac{4}{3} J_{\rm H} {\bf S}_1 . {\bf S}_2 + 2 J_{\rm H} \left [ \mathcal{S}_{1}^z \mathcal{S}_{2}^z - \mathcal{S}_{1}^z \mathcal{S}_{3}^z \right] \nonumber \\
&+& \left( U-\frac{13}{6} J_{\rm H} \right) \left[ n_{1} n_{2} + n_{1} n_{3} \right] + 
\left( U-\frac{7}{3} J_{\rm H} \right) n_{2} n_{3} .
\label{h_int}
\end{eqnarray}
The symmetry features of the interaction terms above are consistent with a general pseudo-spin rotation symmetry analysis which shows that the Hund's coupling ($J_{\rm H}$) and pair-hopping ($J_{\rm H}$) interaction terms in Eq. (\ref{inter}) explicitly break this symmetry systematically, while the Hubbard $(U)$ and density $(U')$ interaction terms do not.\cite{iridate_four} 


\section{Antiferromagnetic state of the three-orbital model}
We consider the various interaction terms in Eq. (\ref{h_int}) in the Hartree-Fock (HF) approximation, focussing on the staggered field terms corresponding to ($\pi,\pi$) ordered AF state on the square lattice. The charge terms corresponding to density condensates will be discussed in the next section. For general ordering direction with components {\boldmath $\Delta_l$}= $(\Delta_l ^x,\Delta_l ^y,\Delta_l ^z)$, the staggered field term for sector $l$ in the pseudo-orbital basis is given by:  
\begin{equation}
\mathcal{H}_{\rm sf} (l) = 
\sum_{{\bf k} s}  \psi_{{\bf k}ls}^{\dagger} 
\begin{pmatrix} -s \makebox{\boldmath $\tau . \Delta_l$}
\end{pmatrix} \psi_{{\bf k}ls} 
= \sum_{{\bf k} s} -s \psi_{{\bf k}ls}^{\dagger} 
\begin{pmatrix} \Delta_l ^z & \Delta_l ^x - i \Delta_l ^y \\
\Delta_l ^x + i \Delta_l ^y & -\Delta_l ^z  \\
\end{pmatrix} \psi_{{\bf k}ls}
\label{gen_ord_dirn} 
\end{equation}  
where $\psi_{{\bf k}ls} ^\dagger = (a_{{\bf k}ls\uparrow} ^\dagger \; \; a_{{\bf k}ls\downarrow} ^\dagger)$, $s=\pm 1$ for the two sublattices A/B, and the staggered field components $\Delta_{l=1,2,3} ^{\alpha=x,y,z}$ are self-consistently determined from:
\begin{eqnarray}
2 \Delta_1 ^\alpha &=& {\mathcal U}_1 m_1 ^\alpha + \frac{2J_{\rm H}}{3} m_2 ^\alpha + J_{\rm H} (m_3 ^\alpha - m_2 ^\alpha) \delta_{\alpha z} \nonumber \\
2 \Delta_2 ^\alpha &=& {\mathcal U}_2 m_2 ^\alpha + \frac{2J_{\rm H}}{3} m_1 ^\alpha - J_{\rm H} m_1 ^\alpha \delta_{\alpha z} \nonumber \\
2 \Delta_3 ^\alpha &=& {\mathcal U}_3 m_3 ^\alpha + J_{\rm H} m_1 ^\alpha \delta_{\alpha z}  
\label{selfcon}
\end{eqnarray}
in terms of the staggered pseudo-spin magnetization components $m_{l=1,2,3} ^{\alpha=x,y,z}$. In practice, it is easier to choose set of ${\bf \Delta}_{l=1,2,3}$ and self-consistently determine the Hubbard-like interaction strengths ${\mathcal U}_{l=1,2,3}$ such that ${\mathcal U}_1 = U - \frac{4}{3} J_{\rm H}$ and ${\mathcal U}_2 = {\mathcal U}_3 =  U - J_{\rm H}$ using Eq. (\ref{selfcon}). 

Transforming the staggered-field term back to the three-orbital basis $(yz\sigma,xz\sigma,xy\bar{\sigma})$, and including the SOC and band terms,\cite{watanabe_PRL_2010} the full HF Hamiltonian considered in our band structure and spin fluctuation analysis is given by
$\mathcal {H}_{\rm HF} = \mathcal{H}_{\rm SO} + \mathcal{H}_{\rm band} + \mathcal{H}_{\rm sf}$,
where,
\begin{eqnarray}
\mathcal{H}_{\rm band} &=& 
\sum_{{\bf k} \sigma s} \psi_{{\bf k} \sigma s}^{\dagger} \left [ \begin{pmatrix}
{\epsilon_{\bf k} ^{yz}}^\prime & 0 & 0 \\
0 & {\epsilon_{\bf k} ^{xz}}^\prime & 0 \\
0 & 0 & {\epsilon_{\bf k} ^{xy}}^\prime \end{pmatrix} \delta_{s s^\prime}
+ 
\begin{pmatrix}
\epsilon_{\bf k} ^{yz} & \epsilon_{\bf k} ^{yz|xz} & 0 \\
-\epsilon_{\bf k} ^{yz|xz} & \epsilon_{\bf k} ^{xz} & 0 \\
0 & 0 & \epsilon_{\bf k} ^{xy} \end{pmatrix} \delta_{\bar{s} s^\prime } \right]
\psi_{{\bf k} \sigma s^\prime} 
\label{three_orb_two_sub}
\end{eqnarray} 
in the composite three-orbital, two-sublattice basis, showing the different hopping terms connecting the same and opposite sublattice(s). Corresponding to the hopping terms in the tight-binding model, the various band dispersion terms in Eq. (\ref{three_orb_two_sub}) are given by: 
\begin{eqnarray}
\epsilon_{\bf k} ^{xy} &=& -2t_1(\cos{k_x} + \cos{k_y}) \nonumber \\
{\epsilon_{\bf k} ^{xy}} ^{\prime} &=& - 4t_2\cos{k_x}\cos{k_y} - \> 2t_3(\cos{2{k_x}} + \cos{2{k_y}}) + \mu_{xy}  \nonumber \\
\epsilon_{\bf k} ^{yz} &=& -2t_5\cos{k_x} -2t_4 \cos{k_y} \nonumber \\
\epsilon_{\bf k} ^{xz} &=& -2t_4\cos{k_x} -2t_5 \cos{k_y}  \nonumber \\
\epsilon_{\bf k} ^{yz|xz} &=&  -2t_{m}(\cos{k_x} + \cos{k_y}) . 
\end{eqnarray}
Here $t_1$, $t_2$, $t_3$ are respectively the first, second, and third neighbor hopping terms for the $xy$ orbital, which has energy offset $\mu_{xy}$ from the degenerate $yz/xz$ orbitals induced by the tetragonal splitting. For the $yz$ ($xz$) orbital, $t_4$ and $t_5$ are the NN hopping terms in $y$ $(x)$ and $x$ $(y)$ directions, respectively. Mixing between $xz$ and $yz$ orbitals is represented by the NN hopping term $t_m$. We have taken values of the tight-binding parameters ($t_1$, $t_2$, $t_3$, $t_4$, $t_5$, $t_{\rm m}$, $\mu_{xy}$, $\lambda$) = (1.0, 0.5, 0.25, 1.028, 0.167, 0.0, -0.7, 1.35) in units of $t_1$, where the energy scale $t_1$ = 280 meV. Using above parameters, the calculated electronic band structure shows AFM insulating state and mixing between pseudo-orbital sectors.\cite{watanabe_PRL_2010,iridate_one} As the pseudo-spin canting is not relevant for the following discussion, we have set $t_{\rm m}$ to zero by going to the locally rotated coordinate frame.


To illustrate the AF state calculation, we have taken staggered field values $\Delta_{l=1,2,3}^x= (0.92,0.08,-0.06)$ in units of $t_1$, which ensures self-consistency for all three orbitals, with the given relations ${\mathcal U}_2$=${\mathcal U}_3$=${\mathcal U}_1$+$J_{\rm H}/3$. Using the calculated sublattice magnetization values $m_{l=1,2,3}^x$=(0.65,0.005,-0.038), we obtain ${\mathcal U}_{l=1,2,3}$=(0.80,0.83,0.83) eV, which finally yields $U$=$\mathcal U_1$+$\frac{4}{3}J_{\rm H}$=0.93 eV for $J_{\rm H}$=0.1 eV. For these parameter values, the calculated magnon dispersion and energy gap are in very good agreement with RIXS measurements.\cite{kim1_PRL_2012,kim_NATCOMM_2014,pincini_PRB_2017,porras_PRB_2019} The easy $x$-$y$ plane anisotropy arising from Hund's coupling results in energy gap $\approx 40$ meV for the out-of-plane ($z$) magnon mode.\cite{iridate_two}  

The electron fillings are obtained as $n_{l=1,2,3}\approx (1.064,1.99,1.946)$ in the three pseudo orbitals. Finite mixing between the $J$=1/2 and 3/2 sectors is reflected in the small deviations from ideal fillings and also in the very small magnetic moment values for $l=2,3$ as given above, which play a crucial role in the expression of true magnetic anisotropy and magnon gap in view of the Hund's coupling induced anisotropic interactions in Eq. (\ref{h_int}). The values $\lambda$=0.38 eV, $U$=0.93 eV, and $J_{\rm H}$=0.1 eV taken above lie well within the estimated parameter range for $\rm Sr_2 Ir O_4$.\cite{zhou_PRX_2017,igarashi_PRB_2014}


\section{Renormalized spin-orbit gap}

In this section, we obtain the relative energy shift between the $J$=1/2 and 3/2 states arising from the transformed density interaction terms in Eq. (\ref{h_int}. This relative shift effectively renormalizes the spin-orbit gap and plays an important role in determining the energy scale of the spin-orbit exciton, as discussed in the next section. Corresponding to the total density condensate $\langle n_{l\uparrow} + n_{l\downarrow} \rangle$ in the HF approximation of the density interaction terms, the spin-independent self-energy contributions for the three pseudo orbitals are obtained as: 
\begin{eqnarray}
\Sigma_{\rm dens}^{l=1} &=& U \left \langle \frac{1}{2}n_1 + n_2 + n_3 \right \rangle - J_{\rm H} \left \langle \frac{2}{3}n_1 + \frac{13}{6} n_2 + \frac{13}{6} n_3 \right \rangle  \nonumber \\
\Sigma_{\rm dens} ^{l=2} &=& U \left \langle n_1 + \frac{1}{2}n_2 + n_3 \right \rangle - J_{\rm H} \left \langle \frac{13}{6}n_1 + \frac{1}{2} n_2 + \frac{7}{3} n_3 \right \rangle \nonumber \\
\Sigma_{\rm dens} ^{l=3} &=& U \left \langle n_1 + n_2 + \frac{1}{2}n_3 \right \rangle - J_{\rm H} \left \langle \frac{13}{6}n_1 + \frac{7}{3} n_2 + \frac{1}{2} n_3 \right \rangle 
\end{eqnarray}
The formally unequal contributions will result in relative energy shifts between the three orbitals depending on the electron filling. With $\langle n_1 \rangle$=1 and $\langle n_2 \rangle$=$\langle n_3 \rangle$=2 for the $d^5$ system having nominally half-filled and filled orbitals, the relative energy shift: 
\begin{equation}
\Delta_{\rm dens} = \Sigma_{\rm dens}^{l=1} - \Sigma_{\rm dens}^{l=2,3} = \frac{U- 3 J_{\rm H}}{2}
\end{equation}
between $l$=1 and (degenerate) $l$=2,3 orbitals. 

For $U > 3J_{\rm H}$, the relative energy shift enhances the energy gap between $J$=1/2 and $3/2$ sectors, effectively resulting in a correlation-induced renormalization of the spin-orbit gap and the spin-orbit coupling. The SOC strength is renormalized as $\tilde{\lambda} = \lambda + 2\Delta_{\rm dens}/3$ by the relative energy shift. With $\Delta_{\rm dens} = (U-3J_{\rm H})/2 \approx$ 0.3 eV for the parameter values considered earlier, we obtain $\tilde{\lambda} \approx$ 0.6 eV, which is in agreement with the correlation-enhanced SOC strength obtained in a recent DFT study of $\rm Sr_2 Ir O_4$.\cite{zhou_PRX_2017} For $d^4$ systems with nominally $\langle n_1 \rangle$=0, the relative energy shift increases to $U-3J_{\rm H}$. This enhancement of the spin-orbit gap renormalization is seen in recent DFT study of the hexagonal iridates $\rm Sr_3 Li Ir O_6$ and $\rm Sr_4 Ir O_6$ with Ir$^{5+}$ ($5d^4$) and Ir$^{4+}$ ($5d^5$) ions, respectively.\cite{ming_PRB_2018} 



\begin{figure}
\vspace*{-30mm}
\hspace*{0mm}
\psfig{figure=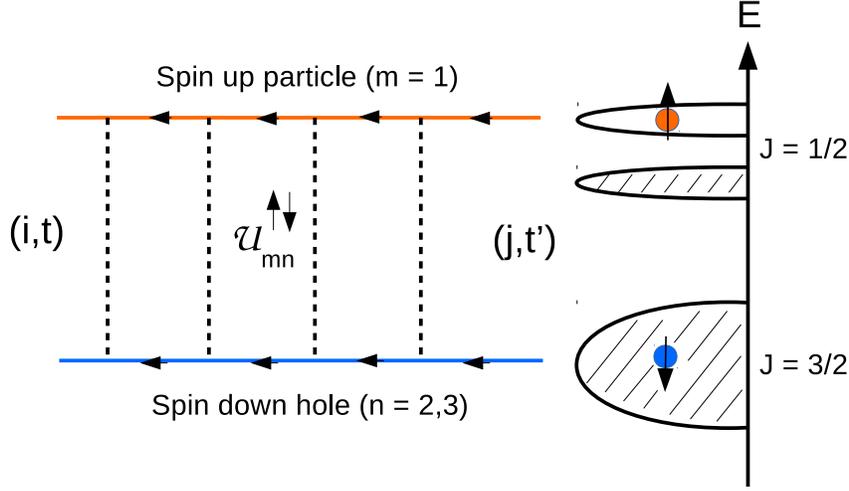,angle=0,width=120mm}
\vspace*{-75mm}
\caption{Propagator of inter-orbital, spin-flip, particle-hole excitations across the renormalized spin-orbit gap between the nominally filled $J$=3/2 sector and the half-filled $J$=1/2 sector.}
\label{fig2}
\end{figure}

\section{Spin-Orbit Exciton}

Magnon excitations modes in $\rm Sr_2 Ir O_4$ essentially involve collective modes of intra-orbital, spin-flip, particle-hole excitations within the magnetically active $J$=1/2 sector.\cite{iridate_one,iridate_two} In analogy, we will investigate here the collective modes of inter-orbital, particle-hole excitations across the renormalized spin-orbit gap between the nominally filled $J$=3/2 sector and the half-filled $J$=1/2 sector. We will consider both pseudo-spin-flip and non-pseudo-spin-flip cases for these spin-orbit exciton modes. Starting first with the spin-flip case, we consider the composite pseudo-spin-orbital fluctuation propagator in the $z$-ordered AFM state:
\begin{equation}
\chi^{-+} _{\rm so} ({\bf q},\omega) = \int dt \sum_{i} e^{i\omega(t-t^\prime)}
e^{-i{\bf q}.({\bf r}_i - {\bf r}_j)}  
\langle \Psi_0 | T [ S_{i,m,n} ^- (t) S_{j,m,n} ^+ (t^\prime) ] | \Psi_0 \rangle
\label{soe_prop}
\end{equation}
involving the inter-orbital spin-raising and -lowering operators $S_{j,m,n}^+$=$a_{jm\uparrow} ^\dagger a_{jn\downarrow}$ and $S_{i,m,n}^-$=$a_{in\downarrow} ^\dagger a_{im\uparrow}$ at lattice sites $j$ and $i$, describing the propagation of a spin-flip, particle-hole excitation between different pseudo orbitals $m$ and $n$. Although the most general propagator would involve $S_{i,m,n}^-$ and $S_{j,m',n'}^+$, the above simplified propagator is a good approximation in view of the orbital restrictions on the particle-hole states as discussed below. Also, we have considered the $z$-ordered AFM state for simplicity as the $J_{\rm H}$-induced weak easy-plane anisotropy has negligible effect on the spin-orbit exciton. 

In the ladder-sum approximation, the spin-orbital propagator is obtained as:
\begin{equation}
[\chi^{-+} _{\rm so} ({\bf q},\omega)] = \frac{[\chi^0 _{\rm so} ({\bf q},\omega)]}
{1 - {\mathcal U} [\chi^0 _{\rm so} ({\bf q},\omega)]}
\label{soe_RPA}  
\end{equation}
where the relevant interactions ${\mathcal U} = \mathcal{U}_{mn}^{\tau \overline{\tau}}$ for the spin-flip particle-hole pair are given in Eq. (\ref{cal_u_eqns}). The ladder-sum approximation with repeated (attractive) interactions (as shown in Fig. (\ref{fig2}) for the retarded case) represents resonant scattering of the particle-hole pair, resulting in a resonant state split-off from the particle-hole continuum, which we identify as the spin-orbit exciton modes. 

The bare particle-hole propagator in the above equation:
\begin{equation}
[\chi^0 _{\rm so} ({\bf q},\omega)]_{s s'} ^{mn} = \sum_{{\bf k}} \left [ 
\frac{\langle \varphi_{\bf k-q} ^n | \tau^- | \varphi_{\bf k} ^m \rangle_s
\langle \varphi_{\bf k} ^m   | \tau^+ | \varphi_{\bf k-q} ^n \rangle_{s'} 
} {E^+_{\bf k-q} - E^-_{\bf k} + \omega - i \eta }
+ \frac{
\langle \varphi_{\bf k-q} ^n | \tau^- | \varphi_{\bf k} ^m \rangle_s
\langle \varphi_{\bf k} ^m   | \tau^+ | \varphi_{\bf k-q} ^n \rangle_{s'} 
} {E^+_{\bf k} - E^-_{\bf k-q} - \omega - i \eta } \right ]
\label{chi0_so}
\end{equation}
was evaluated in the two-sublattice basis by integrating out the fermions in the $(\pi,\pi)$ ordered state. Here $E_{\bf k}$ and $\varphi_{\bf k}$ are the eigenvalues and eigenvectors of the Hamiltonian matrix in the pseudo-spin-orbital basis, and the $E_{\bf k}$ superscript $+(-)$ refers to particle (hole) energies above (below) the Fermi energy. The projected amplitudes $\varphi^m _{{\bf k}\tau}$ above were obtained by projecting the ${\bf k}$ states in the three-orbital basis $|\mu,\sigma\rangle$ on to the pseudo-orbital basis $|m,\tau\rangle$ corresponding to the $J=1/2$ and $3/2$ sector states, as given below:
\begin{eqnarray}
\varphi_{{\bf k}\uparrow}^1 = \frac{1}{\sqrt{3}} \left( \phi^{yz}_{{\bf k}\downarrow} - i\phi^{xz}_{{\bf k}\downarrow} + \phi^{xy}_{{\bf k}\uparrow}\right) \;\;\;\;\;\; & & 
\varphi_{{\bf k}\downarrow}^1 = \frac{1}{\sqrt{3}} \left( \phi^{yz}_{{\bf k}\uparrow} + i \phi^{xz}_{{\bf k}\uparrow} - \phi^{xy}_{{\bf k}\downarrow}\right) \nonumber \\ 
\varphi_{{\bf k}\uparrow}^2 = \frac{1}{\sqrt{6}} \left( \phi^{yz}_{{\bf k}\downarrow} - i\phi^{xz}_{{\bf k}\downarrow} - 2 \phi^{xy}_{{\bf k}\uparrow}\right)  \;\;\;\; & & 
\varphi_{{\bf k}\downarrow}^2 = \frac{1}{\sqrt{6}} \left( \phi^{yz}_{{\bf k}\uparrow} + i \phi^{xz}_{{\bf k}\uparrow} + 2 \phi^{xy}_{{\bf k}\downarrow}\right) \nonumber \\
\varphi_{{\bf k}\uparrow}^3 = \frac{1}{\sqrt{2}} \left( \phi^{yz}_{{\bf k}\downarrow} + i\phi^{xz}_{{\bf k}\downarrow} \right) \;\;\;\;\;\;\;\;\;\;\;\;\;\;\; & & 
\varphi_{{\bf k}\downarrow}^3 = \frac{1}{\sqrt{2}} \left( \phi^{yz}_{{\bf k}\uparrow} - i \phi^{xz}_{{\bf k}\uparrow} \right)
\label{proj_ampl}
\end{eqnarray}
in terms of the amplitudes $\phi^{\mu}_{{\bf k}\sigma}$ in the three-orbital basis $(\mu = yz,xz,xy)$. The $[\chi^0 ({\bf q},\omega)]$ matrix was evaluated by performing the $\bf k$ sum over the 2D Brillouin zone divided into a 300 $\times$ 300 mesh.

The dominant contribution to $[\chi^0 _{\rm so} ({\bf q},\omega)]$ above will correspond to particle $(+)$ states in the nominally half-filled pseudo-orbital $m$=1 ($J$=1/2 sector) and hole $(-)$ states in the nominally filled pseudo-orbitals $n$=2,3 ($J$=3/2 sector). Due to these restrictions, the bare propagator essentially becomes diagonal in the composite particle-hole orbital basis ($m'$=$m$,$n'$=$n$), which justifies the simplified propagator considered above. In order to focus exclusively on the high-energy spin-orbit exciton modes, particle-hole excitations within the $J$=1/2 sector (which yield the low-energy magnon modes) have been excluded. 

\begin{figure}
\vspace*{0mm}
\hspace*{0mm}
\psfig{figure=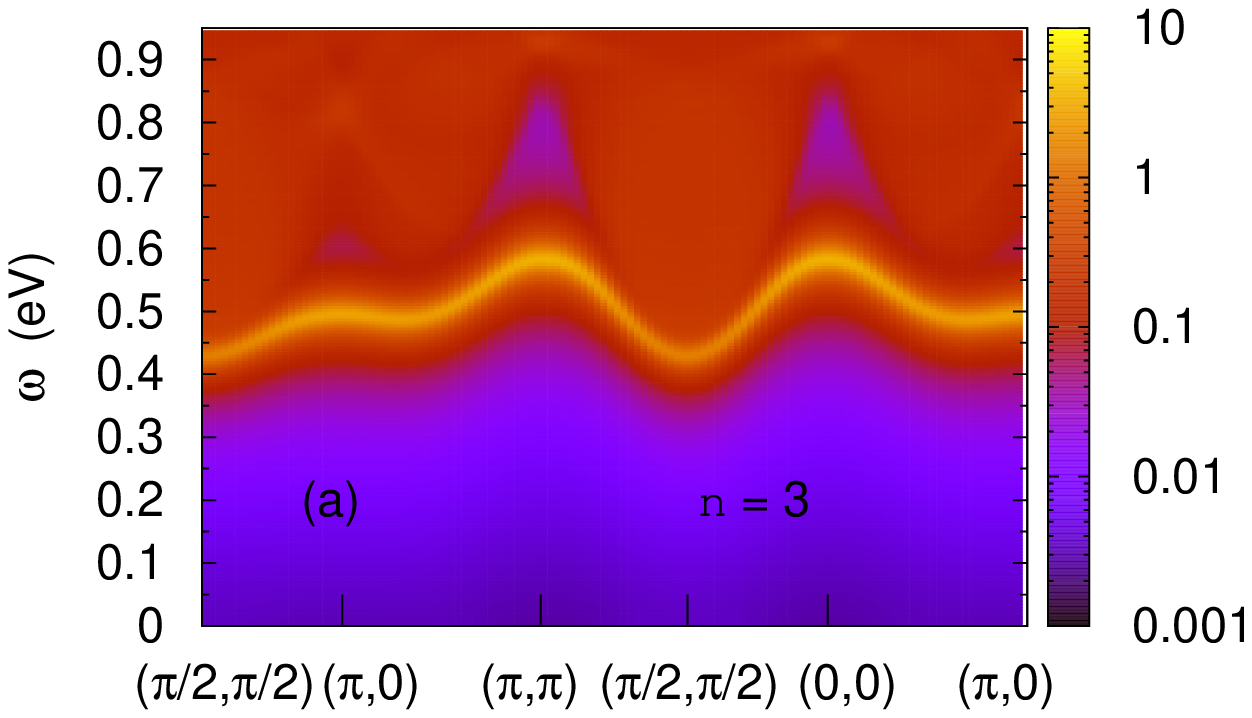,angle=0,width=80mm}
\psfig{figure=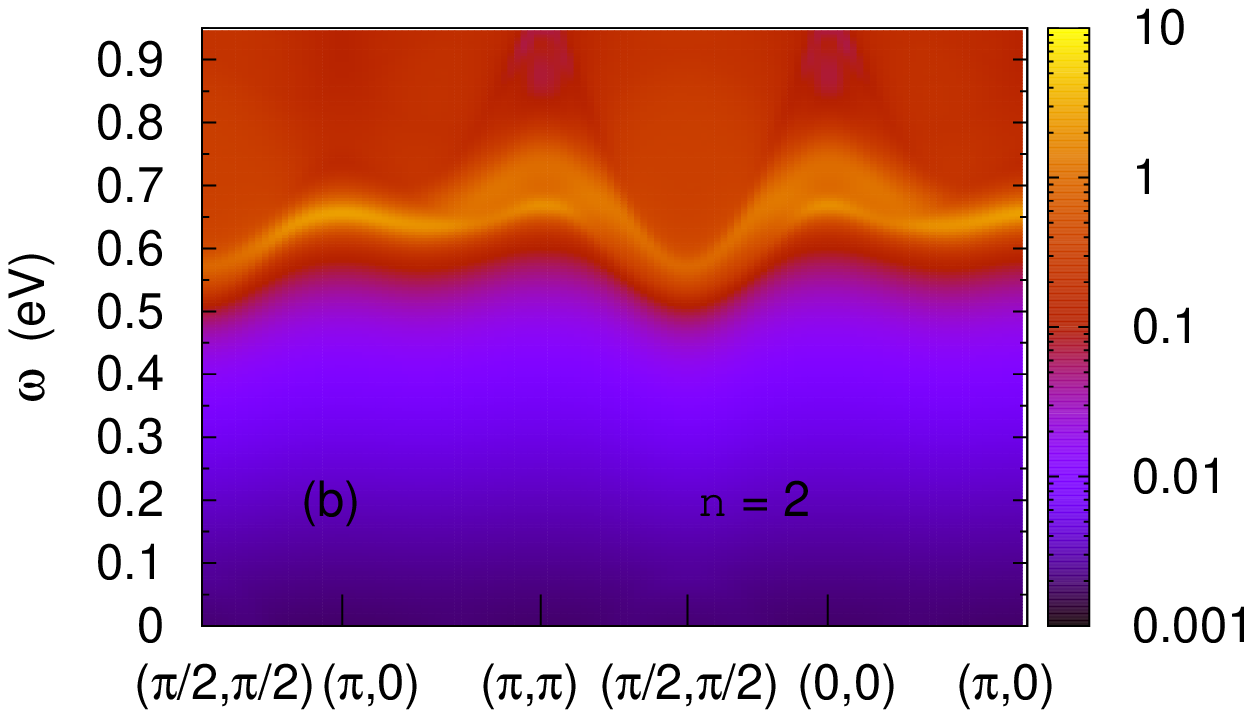,angle=0,width=80mm}
\vspace{-10mm}
\caption{The spin-orbit exciton spectral function $A_{\bf q}(\omega)$ for the two cases: (a) 
$(m,n)$=(1,3) and (b) $(m,n)$=(1,2), showing well defined dispersive modes near the lower edge of the continuum. The exciton represents collective spin-orbital excitations across the renormalized spin-orbit gap.}
\label{spectral}
\end{figure}

Fig. \ref{spectral} shows the spin-orbit exciton spectral function:
\begin{equation}
A_{\bf q}(\omega) = \frac{1}{\pi} {\rm Im \; Tr} \left [ \chi^{-+} _{\rm so} ({\bf q},\omega) \right] 
\end{equation}
as an intensity plot for $\bf q$ along the high symmetry directions of the BZ. For clarity, we have considered here the particle-hole propagator in Eq. (\ref{chi0_so}) separately for $(m,n)$=(1,3) and (1,2). The relevant interaction terms for these two cases are: $\mathcal{U}_{13}^{\tau \overline{\tau}}$=$U$-$5J_{\rm H}/3$ and $\mathcal{U}_{12}^{\tau \overline{\tau}}$=$U$-$7J_{\rm H}/3$. Here, we have taken $U$=0.93 eV and $J_{\rm H}$=0.1 eV as obtained in Sec. III, and the renormalized spin-orbit gap (Sec. IV) has been incorporated. 

The spin-orbit exciton spectral function in Fig. \ref{spectral}(a) clearly shows a well defined propagating mode near the lower edge of the continuum with significantly higher intensity compared to the continuum background. With increasing interaction strength, this mode progressively shifts to lower energy further away from the continuum, and becomes less dispersive and more prominent in intensity, indicating enhanced localization of the spin-orbit exciton. 

Fig. \ref{spectral}(b) shows a similar exciton mode for the other case $(m,n)$=(1,2), with slightly higher energy and reduced dispersion as well as significant damping. The relatively reduced interaction strength $\mathcal{U}_{12}^{\tau \overline{\tau}}$ for this mode accounts for the slightly higher energy. We have similarly obtained the spectral functions for the non-spin-flip cases by considering operators $n_{j,m,n}^\tau$=$a_{jm\tau}^\dagger a_{jn\tau}$ and $n_{i,m,n}^{\tau\dagger}$=$a_{in\tau}^\dagger a_{im\tau}$ instead of $S_{j,m,n}^+$ and $S_{i,m,n}^-$ in Eq. (\ref{soe_prop}) with appropriate interactions $\mathcal{U}_{mn}^{\tau\tau}$. The spectral functions for these cases are nearly identical, as expected from the non-magnetic character of the filled $J$=3/2 sector. 

The calculated dispersion and energy scale of the two spin-orbit exciton modes are in excellent agreement with RIXS measurements in $\rm Sr_2 Ir O_4$.\cite{kim1_PRL_2012,pincini_PRB_2017} Comparison of the calculated $A_{\bf q}(\omega)$ with the observed RIXS intensity and its momentum dependence is beyond the scope of this work. The basic RIXS mechanism involved in the creation of the spin-orbit exciton, whose propagation is considered in Eq. (\ref{soe_RPA}), is explained below. 


The $L_3$-edge RIXS essentially involves second-order dipole-allowed transitions between $2p_{3/2}$ core level and $t_{2g}$ levels. The incoming photon resonantly excites a $2p_{3/2}$ electron to the unfilled $t_{2g}$ states (upper Hubbard band of the nominally $J=1/2$ sector). In the subsequent radiative de-excitation, an electron from the filled $t_{2g}$ states fills the $2p_{3/2}$ core hole, the loss in photon energy thereby corresponding to the overall particle-hole excitation in the $t_{2g}$ manifold. The magnon and spin-orbit exciton cases correspond to the final-state $t_{2g}$ hole created in the $J=1/2$ and 3/2 sectors, respectively.

\begin{figure}
\vspace*{-20mm}
\hspace*{0mm}
\psfig{figure=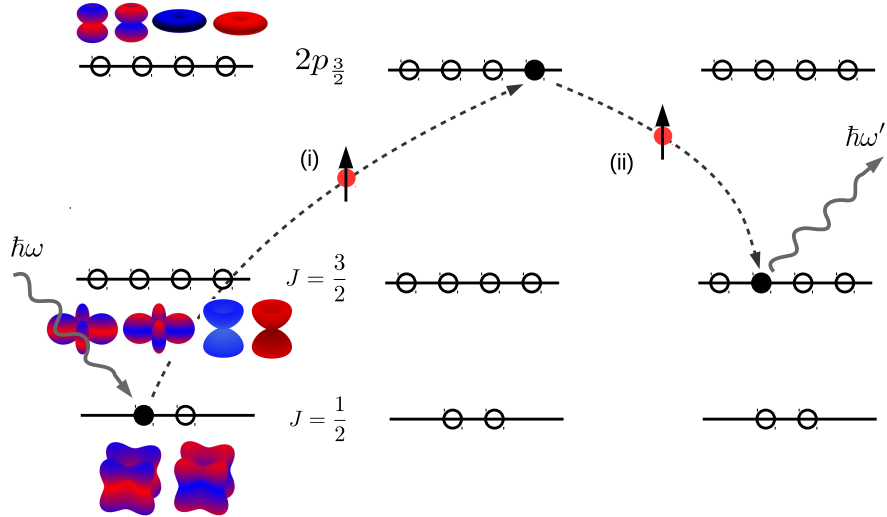,angle=0,width=120mm}
\vspace{0mm}
\caption{The optical (i) excitation ($i \rightarrow 2p_{3/2}$) and (ii) de-excitation ($2p_{3/2} \rightarrow f$) processes (in the hole picture) involved in the RIXS mechanism for the particle-hole excitation across the renormalized spin-orbit gap. The real spin is conserved in optical transitions.}
\label{optical}
\end{figure}

In the magnon case, with both initial and final hole states in the $J=1/2$ sector (in the hole picture), the dipole matrix elements $\langle 2p_{3/2}|{\cal D}_\epsilon|i\rangle$ and $\langle f|{\cal D}^\dagger _{\epsilon '}|2p_{3/2}\rangle$ involving pseudo-spin-flip have been shown to be finite,\cite{ament_PRB_2011,boseggia_2014} implying that RIXS is fully allowed, and the observed low-energy RIXS spectrum corresponds to the magnon excitation. In the spin-orbit exciton case, with final hole state in the $J$=3/2 sector, the optical excitation and de-excitation processes are shown in Fig. \ref{optical}. These processes involve no change in real spin which is conserved in optical transitions.\cite{boseggia_2014} However, due to the spin-orbital entangled nature of the $J$ states, both pseudo-spin-flip and non-pseudo-spin-flip cases are allowed with respect to the initial and final hole states. For example, the pseudo-spin-flip case is realized if $i \rightarrow 2p_{3/2}$ involves excitation of $(xy,\sigma=\uparrow)$ hole from $| l=1,\tau=\uparrow\rangle$ state and $2p_{3/2} \rightarrow f$ involves de-excitation of hole to the $(yz,\sigma=\uparrow)$ component of $|l=3,\tau=\downarrow\rangle$ state. 


\section{Conclusions}


Well-defined propagating spin-orbit exciton modes were obtained representing collective modes of inter-orbital, particle-hole excitations across the renormalized spin-orbit gap, with both dispersion and energy scale in excellent  agreement with RIXS studies. The relevant interaction terms for the two exciton modes as well as the renormalized spin-orbit gap, which play an important role in the spin-orbit exciton energy scale, were obtained from the transformation of the various Coulomb interaction terms to the pseudo-spin-orbital basis formed by the $J$=1/2 and 3/2 states. The approach presented here allows for a unified description of magnons and spin-orbit excitons in spin-orbit coupled systems.

\end{document}